\documentclass[%
 reprint,
 superscriptaddress,
 amsmath,amssymb,
 aps,
 prl
]{revtex4-1}

\usepackage{amsmath}
\usepackage{amsthm}
\usepackage[colorlinks=true,urlcolor=blue,citecolor=blue,linkcolor=blue]{hyperref}
\usepackage[shortlabels]{enumitem}
\usepackage{amsfonts}
\usepackage{graphicx}
\usepackage{bbold}
\usepackage{color}
\usepackage{hyperref}
\usepackage{verbatim}
\usepackage{cases}
\usepackage{amsfonts}
\usepackage{amssymb}
\usepackage[]{mathrsfs}
\usepackage{xcolor}
\usepackage{epsfig}
\usepackage{bm}
\usepackage{setspace}
\usepackage{enumerate}
\usepackage{dcolumn}
\usepackage{bm}
\usepackage{enumitem}

\usepackage{graphicx}
\usepackage{dcolumn}
\usepackage{bm}
\usepackage[FIGTOPCAP,small,tight]{subfigure}
\usepackage{color}

\newcommand{\bra}[1]{\left< #1 \right\vert}
\newcommand{\ket}[1]{\left\vert #1 \right>}
\newcommand{\pare}[1]{\left( #1 \right)}
\newcommand{\abs}[1]{\left\vert #1 \right\vert}
\newcommand{\cor}[1]{\left[ #1 \right]}

\newcommand{\llav}[1]{\left\lbrace #1 \right\rbrace}

\renewcommand{\epsilon}{\varepsilon}

\begin{document}

\preprint{APS/123-QED}

\title{Temperature-Controlled Entangled-Photon Absorption Spectroscopy}

\author{Roberto de J. Le\'on-Montiel}
\email{roberto.leon@nucleares.unam.mx} \affiliation{Instituto de
Ciencias Nucleares, Universidad Nacional Aut\'onoma de M\'exico, Apartado Postal 70-543, 04510 Cd. Mx., M\'exico}
\author{Ji\v{r}\'i Svozil\'{i}k}
\affiliation{Yachay Tech, School of Physical Sciences \&
Nanotechnology, 100119, Urcuqu\'i, Ecuador}
\affiliation{Joint Laboratory of Optics of Palack\'y University
and Institute of Physics of CAS, Faculty of Science, Palack\'y
University, 17. listopadu 12, 771 46 Olomouc, Czech Republic}

\author{Juan P. Torres}
\affiliation{ICFO - Institut de Ciencies Fotoniques, Mediterranean
Technology Park, 08860 Castelldefels (Barcelona), Spain}
\affiliation{Department of Signal Theory and Communications,
Campus Nord D3, Universitat Politecnica de Catalunya, 08034
Barcelona, Spain}
\author{Alfred B. U'Ren}
\affiliation{Instituto de Ciencias Nucleares, Universidad Nacional
Aut\'onoma de M\'exico, Apartado Postal 70-543, 04510 Cd. Mx.,
M\'exico}


\begin{abstract}
Entangled two-photon absorption spectroscopy (TPA) has been widely
recognized as a powerful tool for revealing relevant information
about the structure of complex molecular systems. However, to date, the
experimental implementation of this technique has remained
elusive, mainly because of two major difficulties. First, the need
to perform multiple experiments with two-photon
states bearing different temporal correlations, which translates
in the necessity to have at the experimenter's disposal tens, if not hundreds,
of sources of entangled photons. Second, the need to have \emph{a
priori} knowledge of the absorbing medium's lowest-lying
intermediate energy level. In this work, we put forward a simple
experimental scheme that successfully overcomes these two
limitations. By making use of a temperature-controlled
entangled-photon source, which allows the tuning of the central
frequencies of the absorbed photons, we show that the TPA signal,
measured as a function of the temperature of the nonlinear crystal
that generates the paired photons, and a
controllable delay between them, carries all
information about the electronic level structure of the absorbing
medium, which can be revealed by a simple Fourier transformation.
\end{abstract}

\maketitle


Nonlinear spectroscopy techniques have been used in analytical 
chemistry for extracting information about the dynamics
and structure of complex systems, from small molecules to large light-harvesting photosynthetic complexes \cite{ernst1987,mukamel_book,hamm2011,cho2009,joel_book}. Even though in the optical regime these techniques are typically implemented
using laser light, recent investigations suggest that the use of
nonclassical states of light, such as entangled photon pairs, may open new and
exciting avenues in experimental spectroscopy
\cite{javanainen1990,lee2006,muthukrishnan2004,scully2011,fei1997,guzman2010,saleh1998,kojima2004,perina1998,roberto2013,roslyak2009,roslyak2009-2,raymer2013,schlawin2013,schlawin2016,Schlawin2017,shapiro2011,shapiro_book,Dorfman2016,Oka2010,Villabona2017,Varnavski2017,schlawin2017-njp,Oka2018_1,Oka2018_2,svozilik2018_1,svozilik2018_2,burdick2018}.
Remarkably, quantum light has enabled the observation of fascinating two-photon absorption (TPA) phenomena, such as the
linear dependence of two-photon absorption rate on photon flux
\cite{javanainen1990,lee2006}, as well as the prediction of
intriguing effects, such as inducing disallowed atomic transitions \cite{muthukrishnan2004,scully2011},
two-photon-induced transparency \cite{fei1997,guzman2010}, manipulation of quantum pathways of matter \cite{roslyak2009,roslyak2009-2,raymer2013,schlawin2016,schlawin2013,schlawin2017-njp},
and the control of entanglement in molecular processes \cite{shapiro2011,shapiro_book}.

Among different quantum-enabled techniques, entangled-photon
virtual-state spectroscopy
\cite{saleh1998,kojima2004,perina1998,roberto2013,Dorfman2016,Schlawin2017,Oka2010,Villabona2017,Varnavski2017,Oka2018_1,Oka2018_2,svozilik2018_1,svozilik2018_2,burdick2018} is a promising tool for extracting information about
the intermediate, energy non-conserving electronic transitions
\cite{shore1979,sakurai_book}, that contribute to the two-photon
excitation of a chemical or biological sample. In this technique,
virtual-state transitions, a signature of the absorbing medium,
are experimentally revealed by introducing a time delay between
frequency-correlated photons and averaging over many experimental
realizations with differing two-photon state characteristics \cite{saleh1998}.

Regardless of some experimental concerns already raised in the
original virtual-state spectroscopy (VSS) proposal, VSS has been
broadly considered as a new promising route towards novel
applications in ultra-sensitive detection
\cite{Dorfman2016,Schlawin2018}. In this work, we propose an
experimental scheme that overcomes the two major difficulties that
one encounters when implementing the VSS technique, namely the
need for averaging over experimental realizations differing in
temporal correlations between the photons that are
absorbed---which translates into the need for using hundreds of
nonlinear crystals with different lengths---and the requirement of a
priori knowledge of the absorbing medium's lowest-lying intermediate energy level.

Our scheme makes use of a temperature-controlled entangled-photon
source to show that the TPA signal, to be recorded as a function of the
temperature of the nonlinear crystal that produces the pair of
absorbed photons, and an external signal-idler delay, yields
information about the electronic level structure of the absorbing
medium, which can be revealed by a simple Fourier transformation.
Because of its simplicity, our proposed technique could be realized
using standard and widely used temperature-controlled
entangled-photon sources. In this way, our work opens up a new avenue
towards the first experimental implementation of nonlinear quantum
spectroscopy.


\begin{figure*}[t!]
\begin{center}
\includegraphics[width=18.0cm]{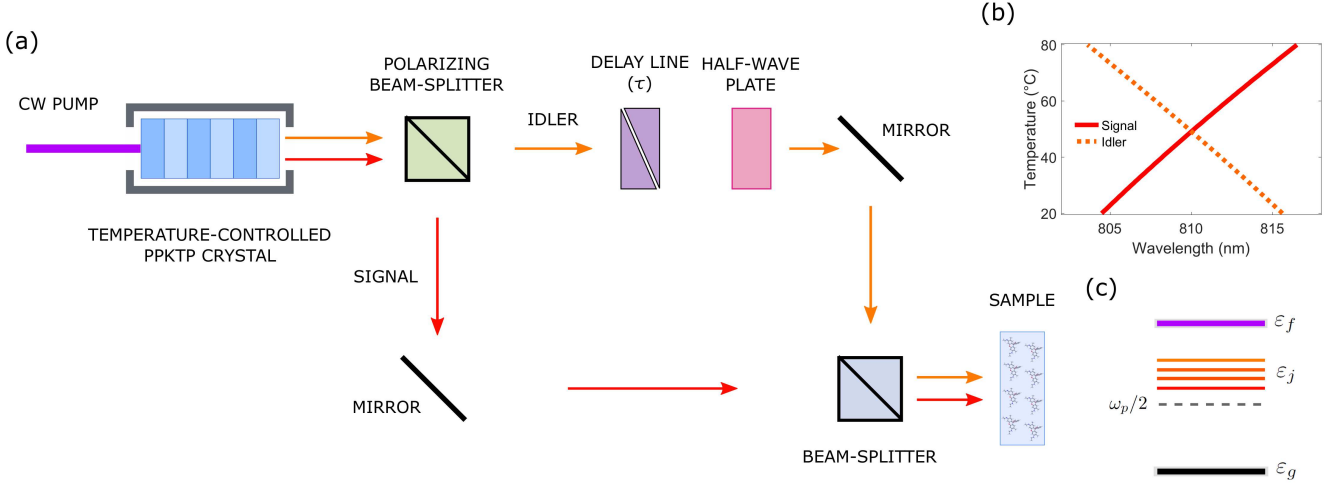}
\end{center}
\vspace{-4mm} \caption{Temperature-controlled quantum nonlinear
spectroscopy. (a) Experimental scheme of the proposed technique.
(b) Central wavelength of the down-converted photons as a function
of the temperature of the PPKTP crystal, considering a continuous
wave pump at $405$ nm. (c) Model of the absorbing medium's
electronic structure.} \label{Fig:setup}
\end{figure*}

In order to describe our proposed spectroscopy technique (shown in Figure
\ref{Fig:setup}), let us consider first the interaction of a
two-photon optical field $\ket{\Psi}$ with a medium described by a
simple energy-level configuration where two-photon transitions
occur from an initial state $\ket{g}$ to a doubly-excited final
state $\ket{f}$ via non-resonant intermediate states denoted by
$\ket{j}$. For simplicity, in the following, we will omit any
other degree of freedom connected to vibrational modes of the
sample, and assume that the lifetimes of intermediate states are
much longer than the light-matter interaction time. Indeed, in
this situation---which can be reached by selecting a proper
correlation time between photons \cite{saleh1998,roberto2013}, the
effects due to dissipation in the single-excitation manifold
(intermediate states) can be considered negligible.

The interaction of an electromagnetic field with a sample, in the
dipole approximation, can be expressed as $\hat{H}\pare{t} =
\hat{d}\pare{t}\hat{E}^{\pare{+}}\pare{t}$, where
$\hat{d}\pare{t}$ stands for the dipole-moment operator and
$\hat{E}^{\pare{+}}\pare{t}$ for the positive-frequency part of
the electric-field operator, which can be expressed as
$\hat{E}^{\pare{+}}\pare{t}=\hat{E}_{1}^{\pare{+}}\pare{t} +
\hat{E}_{2}^{\pare{+}}\pare{t}$. Each of the fields can then be
written as
\begin{equation}
\hat{E}^{\pare{+}}_{1,2}\pare{t} = \int
d\omega_{1,2}\sqrt{\frac{\hbar\omega_{1,2}}{4\pi\epsilon_{0}cA}}\hat{a}_{1,2}\pare{\omega_{1,2}}e^{-i\omega_{1,2}t},
\end{equation}
where $c$ is the speed of light, $\varepsilon_{0}$ is the vacuum
permittivity, $A$ is the effective area of the field interacting
with the sample, and $\hat{a}\pare{\omega_{1,2}}$ is the
annihilation operator of a photonic mode characterized by a
frequency $\omega_{1,2}$ bearing a specific spatial shape and
polarization which, for the sake of simplicity, are not explicitly
written.

By considering that the medium is initially in its ground state
$\ket{g}$ (with energy $\epsilon_{g}$), one can make use of
second-order time-dependent perturbation theory to find that the
probability that the medium is excited to the final state
$\ket{f}$ (with energy $\epsilon_{f}$), through a TPA process, is
given by \cite{perina1998,roberto2013}
\begin{equation}
P_{g \rightarrow f} =
\abs{\frac{1}{\hbar^{2}}\int_{-\infty}^{\infty}dt_{2}\int_{-\infty}^{t_{2}}dt_{1}M_{\hat{d}}\pare{t_1,t_2}M_{\hat{E}}\pare{t_1,t_2}}^{2},
\end{equation}
with
\begin{eqnarray}
M_{\hat{d}}\pare{t_1,t_2} &=& \sum_{j=1} D^{\pare{j}}e^{-i\pare{\epsilon_{j}-\epsilon_{f}}t_2}e^{-i\pare{\epsilon_{g}-\epsilon_{j}}t_1}, \label{Eq:dipole}\\
M_{\hat{E}}\pare{t_1,t_2} &=&
\bra{\Psi_{f}}\hat{E}_{2}^{+}\pare{t_2}\hat{E}_{1}^{+}\pare{t_1}\ket{\Psi}
\nonumber \\ & & \hspace{5mm} +
\bra{\Psi_{f}}\hat{E}_{1}^{+}\pare{t_2}\hat{E}_{2}^{+}\pare{t_1}\ket{\Psi},
\label{Eq:field}
\end{eqnarray}
where $D^{\pare{j}}=\bra{f}\hat{d}\ket{j}\bra{j}\hat{d}\ket{g}$
are the transition matrix elements of the dipole-moment operator.
Note from Eq. (\ref{Eq:dipole}) that excitation of the medium
proceeds through the intermediate states $\ket{j}$, with energy
eigenvalues $\epsilon_{j}$. Also, note that in Eq.
(\ref{Eq:field}), we have only written the terms in which one
photon from each field contributes to the TPA process. In Eq.
(\ref{Eq:field}), $\ket{\Psi_{f}}$ denotes the final state of the
optical field, which we take to be the vacuum state.



\begin{figure}[t!]
\begin{center}
\includegraphics[width=8.75cm]{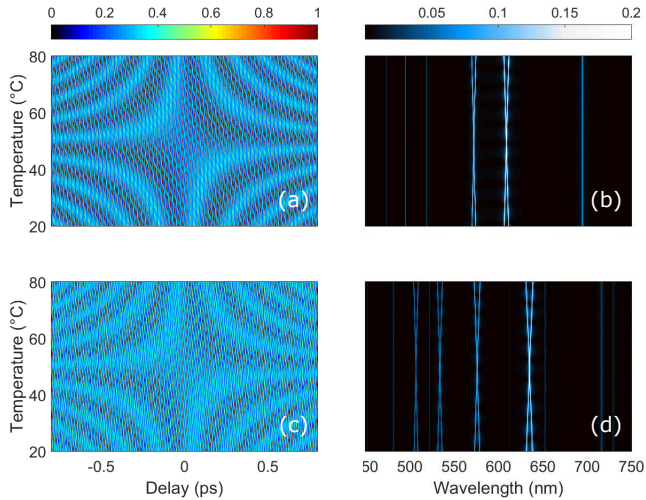}
\end{center}
\vspace{-2mm}
\caption{Entangled-photon absorption spectroscopy for a system
with two-photon transitions taking place via two [(a) and (b)] and
four intermediate states [(c) and (d)]. (a) and (c): Normalized
TPA signal as a function of the crystal temperature $\text{T}$
and the external delay $\tau$. (b) and (d): Normalized Fourier
transform of the TPA signal with respect to the delay. We
assume a source of entangled photon pairs with an entanglement time of
$T_e = 0.87$ ps. For the absorbing medium, the energies of the
levels (or wavelengths) are randomly chosen to be $\lambda_{j}^{(2)}
\in \llav{563, 612}$ nm, and $\lambda_{j}^{(4)} \in \llav{507, 534,
576, 635}$ nm, respectively. Finally, in order to obtain the precise location
of the intermediate states, the frequency axis is displaced by the
degenerate frequency $\omega_0 = 2\pi c/(810$nm) [see supplementary
materials for further details].} \label{Fig:VSS}
\end{figure}

For the sake of simplicity and so as to show the technological readiness
of our proposal, we consider a source of entangled two-photon
states commonly found in many laboratories [see Figure
\ref{Fig:setup}(a)]. Collinear photon pairs with orthogonal
polarizations are generated via type-II spontaneous parametric
down-conversion (SPDC) in a periodically poled $\text{KTiOPO}_{4}$
(PPKTP) crystal of length $L$. We assume continuous wave pumping
of the crystal at $405$ nm. This configuration causes frequency
anti-correlation of the down-converted photons and provides the
strongest TPA signal (see supplementary materials and Refs.
\cite{roberto2013,svozilik2018_1,svozilik2018_2}). The wavelengths
of the photons can be tuned around the degenerate wavelength
($810$ nm) changing the temperature $\text{T}$ of the crystal
\cite{fedrizzi2007}. The generated two-photon state can then be
written as \cite{fedrizzi2009}
\begin{eqnarray}
\ket{\Psi} &=& \pare{\frac{T_{e}}{\sqrt{\pi}}}^{1/2}\int_{-\infty}^{\infty}\int_{-\infty}^{\infty} d\omega_s d\omega_i \delta\pare{\omega_p-\omega_s-\omega_i} \nonumber \\
& & \times
\text{sinc}\llav{T_{e}\cor{\nu-\mu\pare{\text{T}}}}e^{i\omega_{i}\tau}\hat{a}^{\dagger}_{s}\pare{\omega_{s}}\hat{a}^{\dagger}_{i}\pare{\omega_{i}}\ket{0},
\label{Eq:two-photon}
\end{eqnarray}
where $\nu=\omega_{i}-\omega_{s}$, with $\omega_{j}$ ($j=p,s,i$)
representing the frequencies of the pump, signal, and idler
fields, respectively. The correlation (entanglement) time between
the down-converted photons is given by $T_{e}=\pare{N_s-N_i}L/4$,
$N_{s,i}$ being the inverse group velocities of the signal and
idler photons, respectively. In writing Eq. (\ref{Eq:two-photon})
we have assumed that an external delay $\tau$ between the
generated photons has been introduced. As illustrated in Fig.
1(a), this can be experimentally implemented by introducing an
optical delay line for one of the photons, once the pair has been split
by means of a polarizing beam-splitter (PBS).
Note that an additional half-wave plate is used in order to
guarantee that both photons have the same polarization when
impinging on the sample. The non-degeneracy of the photon
wavelengths is given by the function
$\mu\pare{\text{T}}=\omega_{i}^{0}\pare{\text{T}} -
\omega_{s}^{0}\pare{\text{T}}$, where
$\omega_{i,s}^{0}\pare{\text{T}}$ stands for the
temperature-dependent central frequencies of each of the photon
wavepackets. Figure \ref{Fig:setup}(b) shows the dependence of the
down-converted photon central wavelengths on the temperature of
the PPKTP crystal.


The proposed quantum spectroscopy protocol works as follows. We
consider the model system shown in Figure \ref{Fig:setup}(c), in
which the two-photon excitation energy of the medium
$\ket{g}\rightarrow\ket{f}$ corresponds to the pump wavelength
$\lambda_{p} = 405$ nm. By substituting Eqs. (3)-(5) into Eq. (2),
one finds that the TPA signal is expressed as
\begin{widetext}
\begin{equation}
P_{g\rightarrow f}\pare{\tau,\text{T}} =
\frac{\abs{\delta\pare{\frac{\Delta_{+}}{2\pi}}}^2}{4\pi\hbar^2\varepsilon_{0}^2c^2A^2}\frac{\omega_i^{0}\pare{\text{T}}\omega_s^{0}\pare{\text{T}}}{T_{e}}
\abs{ \sum_{j=1}D^{\pare{j}}\llav{ \frac{1 -
e^{-\text{i}\cor{\epsilon_{j} -
\omega_{i}^{0}\pare{\text{T}}}\pare{2T_{e}-\tau}}
}{\epsilon_{j}-\omega_{i}^{0}\pare{\text{T}}} + \frac{1 -
e^{-\text{i}\cor{\epsilon_{j} -
\omega_{s}^{0}\pare{\text{T}}}\pare{2T_{e}+\tau}
}}{\epsilon_{j}-\omega_{s}^{0}\pare{\text{T}}}} }^{2},
\end{equation}
\end{widetext}
where $\Delta_{+} = \pare{\omega_{p}-\epsilon_f}/2$. For the sake
of simplicity, we have displaced our energy levels so that
$\epsilon_{g} = 0$. Furthermore, we have assumed the condition
$\omega_{s}^{0}\pare{\text{T}} + \omega_{i}^{0}\pare{\text{T}} =
\epsilon_{f}$, which guarantees that the
two-photon field is resonant with the transition from the ground
state $\ket{g}$ to the final state $\ket{f}$.

In order to show the usefulness of the technique
proposed for revealing the electronic structure of a molecule,
Figure 2 shows results for two examples. Figures 2(a) and (b)
correspond to a system with two intermediate-state levels whose
energies (in terms of wavelength) are randomly chosen to be
$\lambda_{j}^{(2)} \in \llav{563, 612}$ nm. Figures 2(c) and (d)
correspond to another system with four intermediate-state levels
arbitrarily chosen to be $\lambda_{j}^{(4)} \in \llav{507, 534, 576, 635}$ nm.
Figures 2(a) and (c) show the normalized TPA signals as a function of the crystal's
temperature and the external delay between photons. The
non-monotonic behavior of the TPA signal results from the
interference between different pathways through which two-photon
excitation of the medium occurs \cite{saleh1998,perina1998} and,
more importantly, it shows that the absorption properties of the
sample can be tuned by properly controlling the time and frequency
properties of the entangled photons \cite{Dorfman2016}.

Figures 2(b) and (d) show the normalized Fourier transform of the TPA signals
with respect to the external delay. Surprisingly, two
characteristic patterns of X-shaped and straight lines appear. The
X-shaped lines indicate the energy of intermediate states
($\epsilon_j$), whereas the straight lines appear at the combined
frequencies $\pm\cor{\epsilon_j\pm\epsilon_k}$. The reason behind
this contrasting behavior lies in the fact that the TPA signal of
the former contain frequency components that are temperature
dependent, while the latter are constant with temperature (see
calculations in the supplementary materials for a detailed
analysis on the characteristics of the TPA signal's Fourier
transform).

Besides the remarkable advantage of not requiring many different
sources of entangled photons in the implementation of two-photon
absorption spectroscopy, our scheme permits the direct and straightforward
extraction of the electronic structure of an arbitrary
sample by simply identifying the X-shaped lines, without resorting
to many-sample averaging or sophisticated data analysis. This is
indeed a notable feature that previous proposals have failed to
provide.

In conclusion, we have proposed an experimentally-feasible, novel scheme for absorption spectroscopy based on non-classical light. Previous proposals required the use of multiple nonlinear optics sources, which made its experimental implementation an unrealistic endeavor. Moreover, the acquisition of useful information from the measured data was not straightforward, calling for sophisticated data analysis in order
to isolate information about the sought-after energy level
structure of the sample under investigation. All of this has
prevented the use and further development of entangled-photon
virtual-state spectroscopy.

Contrary to all previous schemes, our proposal makes use of a {\em single} temperature-controlled nonlinear crystal, a state-of-the-art technology widely used
nowadays. Furthermore, the information about the energy level structure can be obtained directly from the experimental data. This indeed constitutes a major simplification of the technique and establishes a new route towards its first experimental demonstration, a milestone in the development of a unique tool capable of providing new and detailed information about the dynamics and chemical structure of complex molecular systems.

\section{Acknowledgments}

We acknowledge Jan Pe\v{r}ina Jr. for helpful discussions. We
thank Javier A. L\'{o}pez Alfaro for his help during the initial
stage of the project, and Mario A. Quiroz-Ju\'{a}rez for his help
in designing the schematic representation of the proposed
technique. This work was supported by DGAPA-UNAM under the project
UNAM-PAPIIT IA100718, and by CONACYT under the project
CB-2016-01/284372. JS thanks the project No. 17-23005Y of Czech Science Foundation. JPT acknowledges financial support from Fundacio Cellex, from the Government of Spain through the Severo Ochoa Programme for Centres of Excellence in R\&D (SEV-2015-0522), from Generalitat de Catalunya under the programs ICREA Academia and CERCA, and from the project 17FUN01 BeCOMe within the Programme EMPIR, and initiative co-founded by the European Union and the EMPIR Participating Countries. AU acknowledges support from PAPIIT (UNAM) grant IN104418, CONACYT Fronteras de la Ciencia grant 1667, and AFOSR grant FA9550-16-1-1458.

\clearpage
\appendix
\setcounter{equation}{0}
\renewcommand{\theequation}{S.\arabic{equation}}
\begin{widetext}

\begin{flushleft}
\singlespacing{\Large{\textbf{Supplementary material:\vspace{2mm}\\
\Large{Temperature-Controlled Entangled-Photon Absorption
Spectroscopy}}}}
\end{flushleft}
\vspace{0.5cm}

\noindent In this supplementary material we
present: (i) the explicit derivation of the different
contributions arising from the Fourier transform of the two-photon
absorption (TPA) signal that would be obtained in the experimental
implementation of temperature-controlled entangled-photon
absorption spectroscopy; and (ii) the relationship between the
strength of the TPA signal and the spectral shape of the absorbed
photons.

\begin{flushleft}
\large{\textbf{1. Temperature-controlled two-photon absorption}}
\end{flushleft}

Let us start by writing the TPA probability as \cite{roberto2013}
\begin{equation}
P_{g \rightarrow f} =
\abs{\frac{1}{\hbar^{2}}\int_{-\infty}^{\infty}dt_{2}\int_{-\infty}^{t_{2}}dt_{1}M_{\hat{d}}\pare{t_1,t_2}M_{\hat{E}}\pare{t_1,t_2}}^{2},
\end{equation}
with
\begin{eqnarray}
M_{\hat{d}}\pare{t_1,t_2} &=& \sum_{j=1} D^{\pare{j}}e^{-i\pare{\epsilon_{j}-\epsilon_{f}}t_2}e^{-i\pare{\epsilon_{g}-\epsilon_{j}}t_1}, \\
M_{\hat{E}}\pare{t_1,t_2} &=&
\bra{\Psi_{f}}\hat{E}_{2}^{+}\pare{t_2}\hat{E}_{1}^{+}\pare{t_1}\ket{\Psi}+
\bra{\Psi_{f}}\hat{E}_{1}^{+}\pare{t_2}\hat{E}_{2}^{+}\pare{t_1}\ket{\Psi},
\end{eqnarray}
where $D^{\pare{j}}=\bra{f}\hat{d}\ket{j}\bra{j}\hat{d}\ket{g}$
are the transition matrix elements of the dipole-moment operator,
and $\epsilon_{g,j,f}$ are the energies of the ground,
intermediate and final states, respectively. The states
$\ket{\Psi}$ and $\ket{\Psi_{f}}$ are the initial and final states
of the optical field. Note that, because we are interested in
the absorption of two-photon states, the final state of the field
is taken to be the vacuum. For the positive part of the electric
field we write
\begin{equation}
\hat{E}^{\pare{+}}_{1,2}\pare{t} = \int
d\omega_{1,2}\sqrt{\frac{\hbar\omega_{1,2}}{4\pi\epsilon_{0}cA}}\hat{a}_{1,2}\pare{\omega_{1,2}}e^{-i\omega_{1,2}t},
\end{equation}
where $c$ is the speed of light, $\varepsilon_{0}$ is the vacuum
permittivity, $A$ is the effective area of the field interacting
with the sample, and $\hat{a}\pare{\omega_{1,2}}$ is the
annihilation operator of a photonic mode with frequency
$\omega_{1,2}$.

Our model is then complete by defining the temperature-controlled
entangled-photon state that interacts with the sample, as
\cite{fedrizzi2009}
\begin{eqnarray}
\ket{\Psi} &=&
\pare{\frac{T_{e}}{\sqrt{\pi}}}^{1/2}\int_{-\infty}^{\infty}\int_{-\infty}^{\infty}
d\omega_s d\omega_i \delta\pare{\omega_p-\omega_s-\omega_i}
\text{sinc}\llav{T_{e}\cor{\nu-\mu\pare{\text{T}}}}e^{i\omega_{i}\tau}\hat{a}^{\dagger}_{s}\pare{\omega_{s}}\hat{a}^{\dagger}_{i}\pare{\omega_{i}}\ket{0},
\end{eqnarray}
where $\nu=\omega_{i}-\omega_{s}$, with $\omega_{j}$ ($j=p,s,i$)
representing the frequencies of the pump, signal, and idler
fields, respectively. The correlation time between the
down-converted photons is given by $T_{e}=\pare{N_s-N_i}L/4$,
$N_{s,i}$ being the inverse group velocities of the signal and
idler photons, respectively. $\tau$ is an external delay between
the generated photons, and their temperature-controlled non-degeneracy
is given by the function
$\mu\pare{\text{T}}=\omega_{i}^{0}\pare{\text{T}} -
\omega_{s}^{0}\pare{\text{T}}$, where
$\omega_{i,s}^{0}\pare{\text{T}}$ are the temperature-dependent
central frequencies of the photons.

By substituting Eqs. (S.2)-(S.5) into Eq. (S.1), it is straightforward
to find that the temperature-dependent TPA signal is given by
\begin{equation}
P_{g\rightarrow f} =
\frac{\abs{\delta\pare{\frac{\Delta_{+}}{2\pi}}}^2}{4\pi\hbar^2\varepsilon_{0}^2c^2A^2}\frac{\omega_i^{0}\pare{\text{T}}\omega_s^{0}\pare{\text{T}}}{T_{e}}
\abs{ \sum_{j=1}D^{\pare{j}}\llav{ \frac{1 -
e^{-\text{i}\cor{\epsilon_{j} -
\omega_{i}^{0}\pare{\text{T}}}\pare{2T_{e}-\tau}}
}{\epsilon_{j}-\omega_{i}^{0}\pare{\text{T}}} + \frac{1 -
e^{-\text{i}\cor{\epsilon_{j} -
\omega_{s}^{0}\pare{\text{T}}}\pare{2T_{e}+\tau}
}}{\epsilon_{j}-\omega_{s}^{0}\pare{\text{T}}}} }^{2},
\end{equation}
where $\Delta_{+} = \pare{\omega_{p}-\epsilon_f}/2$. For the sake
of simplicity, we have displaced our energy levels, so that
$\epsilon_{g} = 0$. Furthermore, we have assumed the condition
$\omega_{s}^{0}\pare{\text{T}} + \omega_{i}^{0}\pare{\text{T}} =
\epsilon_{f}$, which guarantees that the
two-photon field is resonant with the two-photon induced atomic
transition.

So as to understand the behavior of the Fourier transform of the TPA
probability, in particular the ``X'' shape at the
intermediate-level energy positions, we first realize that the
photon spectral non-degeneracy as a function of the crystal's temperature
[depicted in Fig. 1(b) of the main manuscript] can be described by
\begin{eqnarray}
\omega_{s}^{0}\pare{\text{T}} &=& \omega_{0} + \Delta\pare{\text{T}},\\
\omega_{i}^{0}\pare{\text{T}} &=& \omega_{0} -
\Delta\pare{\text{T}},
\end{eqnarray}
where $\omega_0$ is the degenerate frequency, $\omega_0 = 2\pi c/(810$nm)
for our source, and $\Delta\pare{\text{T}}$ is a
temperature-dependent frequency shift.

\noindent By substituting Eqs. (S.7)-(S.8) into
Eq. (S.6), one finds that the TPA signal
can be written explicitly as
\begin{eqnarray}
P_{g\rightarrow f} &=& \sum_{j,k=1}\Bigg\{ \cor{ \frac{1}{\epsilon_{j}-\omega_{0}-\Delta\pare{\text{T}}} + \frac{1}{\epsilon_{j}-\omega_{0}+\Delta\pare{\text{T}}} }\cor{ \frac{1}{\epsilon_{k}-\omega_{0}-\Delta\pare{\text{T}}} + \frac{1}{\epsilon_{k}-\omega_{0}+\Delta\pare{\text{T}}} } \nonumber \\
&-& \cor{ \frac{1}{\epsilon_{j}-\omega_{0}-\Delta\pare{\text{T}}} + \frac{1}{\epsilon_{j}-\omega_{0}+\Delta\pare{\text{T}}} } \cor{ \frac{e^{\text{i}\cor{\epsilon_k-\omega_0-\Delta\pare{\text{T}}}\pare{2T_{e}-\tau}}}{\epsilon_{k}-\omega_{0}-\Delta\pare{\text{T}}} + \frac{e^{\text{i}\cor{\epsilon_k-\omega_0+\Delta\pare{\text{T}}}\pare{2T_{e}+\tau}}}{\epsilon_{k}-\omega_{0}+\Delta\pare{\text{T}}} } \nonumber \\
&-& \cor{ \frac{e^{-\text{i}\cor{\epsilon_j-\omega_0-\Delta\pare{\text{T}}}\pare{2T_{e}-\tau}}}{\epsilon_{j}-\omega_{0}-\Delta\pare{\text{T}}} + \frac{e^{-\text{i}\cor{\epsilon_j-\omega_0+\Delta\pare{\text{T}}}\pare{2T_{e}+\tau}}}{\epsilon_{j}-\omega_{0}+\Delta\pare{\text{T}}} } \cor{ \frac{1}{\epsilon_{k}-\omega_{0}-\Delta\pare{\text{T}}} + \frac{1}{\epsilon_{k}-\omega_{0}+\Delta\pare{\text{T}}} } \nonumber \\
&+& \frac{e^{-\text{i}\cor{\epsilon_j-\epsilon_k}\pare{2T_e-\tau}}}{\cor{\epsilon_j-\omega_0-\Delta\pare{\text{T}}}\cor{\epsilon_k-\omega_0-\Delta\pare{\text{T}}}} + \frac{e^{-\text{i}\cor{\epsilon_j-\epsilon_k}\pare{2T_e+\tau}}}{\cor{\epsilon_j-\omega_0+\Delta\pare{\text{T}}}\cor{\epsilon_k-\omega_0+\Delta\pare{\text{T}}}} \nonumber \\
&+&
\frac{e^{-\text{i}\cor{\epsilon_j-\epsilon_k+2\Delta\pare{\text{T}}}2T_e}e^{-\text{i}\pare{\epsilon_j+\epsilon_k-2\omega_0}\tau}}{\cor{\epsilon_j-\omega_0+\Delta\pare{\text{T}}}\cor{\epsilon_j-\omega_0-\Delta\pare{\text{T}}}}
+
\frac{e^{-\text{i}\cor{\epsilon_j-\epsilon_k-2\Delta\pare{\text{T}}}2T_e}e^{\text{i}\pare{\epsilon_j+\epsilon_k-2\omega_0}\tau}}{\cor{\epsilon_j-\omega_0-\Delta\pare{\text{T}}}\cor{\epsilon_j-\omega_0+\Delta\pare{\text{T}}}}.
\end{eqnarray}

We are now ready to discuss the signals appearing in the Fourier
transform of the TPA signal. Firstly, we realize that the terms
that contribute to the Fourier transform of Eq. (S.9)
are only those depending on the delay $\tau$. Therefore, the first
line represents a constant contribution centered at zero
frequency; the second and third lines produce a signal centered at
the frequencies
$\pm\cor{\tilde{\epsilon}_j\pm\Delta\pare{\text{T}}}$, with
$\tilde{\epsilon}_j = \epsilon_j-\omega_0$. Note that when the
photons are degenerate, i.e. $\Delta\pare{\text{T}}=0$, the
signals appear at the position of the intermediate-state energy
values. Remarkably, as the temperature is increased (or reduced),
the signal splits into two new ones separated by an energy gap
equal to $2\Delta\pare{\text{T}}$, thus producing a frequency plot
similar to the one shown in Fig. 1(b) of the main text. This is the
reason for the ``X'' shape of the signals centered at the
intermediate-state energies. Finally, the fourth and fifth lines
produce signals centered at the frequencies
$\pm\cor{\epsilon_j\pm\epsilon_k}$, which are independent of the
temperature. These contributions are the straight, constant lines
that accompany the intermediate-state X-shaped signals.

\begin{flushleft}
\large{\textbf{2. Role of photon spectra in two-photon absorption spectroscopy}}
\end{flushleft}

In the main text we highlight the fact that the TPA signal shows
its maximum amplitude when the absorbed photons are frequency
anti-correlated. Although this has been extensively discussed in
Refs. \cite{roberto2013,svozilik2018_1,svozilik2018_2}, it is
worth describing the reason behind this important result.

In order to obtain a general form of the TPA signal, we assume that the
nonlinear crystal is now pumped by a Gaussian pulse with temporal
duration $T_{p}$. In this situation, the two-photon state can be
written as \cite{roberto2013}
\begin{eqnarray}
\ket{\Psi} &=& \pare{\frac{T_{p}T_{e}}{2\pi\sqrt{\pi}}}^{1/2}\int_{-\infty}^{\infty}\int_{-\infty}^{\infty} d\omega_s d\omega_i \exp\cor{-T_{p}^{2}\pare{\omega_p-\omega_s-\omega_i}^2} \text{sinc}\llav{T_{e}\cor{\nu-\mu\pare{\text{T}}}} \nonumber \\
& & \hspace{38mm} \times
e^{i\omega_{i}\tau}\hat{a}^{\dagger}_{s}\pare{\omega_{s}}\hat{a}^{\dagger}_{i}\pare{\omega_{i}}\ket{0}.
\end{eqnarray}
The frequency correlations of the down-converted photons can be
tuned by carefully selecting the values of $T_p$ and $T_e$.
Figures 3(a)-(c) show the joint probability distribution of the
two-photon state, i.e.
\begin{equation}
S\pare{\omega_s,\omega_i} =
\frac{T_{p}T_{e}}{2\pi\sqrt{\pi}}\exp\cor{-2T_{p}^{2}\pare{\omega_p-\omega_s-\omega_i}^2}
\text{sinc}^{2}\llav{T_{e}\cor{\nu-\mu\pare{\text{T}}}},
\end{equation}
which represents the probability of detecting a signal photon, with
frequency $\omega_s$, in coincidence with an idler photon, of
frequency $\omega_i$. Frequency anti-correlated photons [Fig.
3(a)] are obtained when $T_p \gg T_e $, whereas for $T_p \ll T_e$
we obtain frequency-correlated photons [Fig. 3(c)]. In the
particular case when $T_p = T_e/2$, frequency quasi-uncorrelated
pairs of photons [Fig. 3(b)] are generated.

\begin{figure}[t!]
\begin{center}
\includegraphics[width=15.5cm]{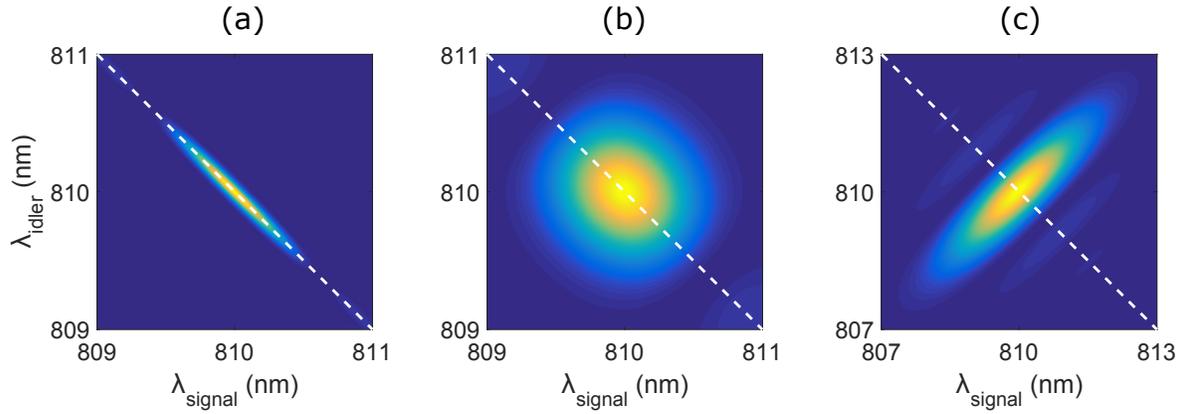}
\end{center}
\caption{Joint spectrum of the two-photon state for different
values of $T_{p}$: (a) $T_{p}=4$ ps, (b) $T_{p}=435$ fs, and (c)
$T_{p}=100$ fs. In all cases, the entanglement time is set to
$T_{e}=0.87$ ps. The dashed line represents the final-state
two-photon resonance condition. Note that the largest overlap
between the two-photon fields and the final state transition
occurs for frequency anti-correlated photons.} \label{Fig:VSS}
\end{figure}

By making use of Eq. (S.10), it is straightforward to find that the
corresponding TPA signal is given by
\begin{eqnarray}
P_{g\rightarrow f} &=& \frac{T_{p}}{T_{e}}\frac{\sqrt{2\pi}\omega_i^{0}\pare{\text{T}}\omega_s^{0}\pare{\text{T}}}{\hbar^2\varepsilon_{0}^2c^2A^2}\exp\cor{-2T_{p}^{2}\pare{\omega_p-\omega_s-\omega_i}} \nonumber \\
&& \hspace{10mm} \times \abs{ \sum_{j=1}D^{\pare{j}}\llav{ \frac{1
- e^{-\text{i}\cor{\epsilon_{j} -
\omega_{i}^{0}\pare{\text{T}}}\pare{2T_{e}-\tau}}
}{\epsilon_{j}-\omega_{i}^{0}\pare{\text{T}}} + \frac{1 -
e^{-\text{i}\cor{\epsilon_{j} -
\omega_{s}^{0}\pare{\text{T}}}\pare{2T_{e}+\tau}
}}{\epsilon_{j}-\omega_{s}^{0}\pare{\text{T}}}} }^{2}.
\end{eqnarray}

Note that the strength of the TPA signal is essentially
controlled by the ratio $T_{p}/T_{e}$, which maximizes when
$T_{p}\gg T_{e}$, i.e., when the photons are frequency
anti-correlated. Physically, this important result can be
understood in terms of the overlap between the joint spectrum of
the entangled photons and the two-photon resonance condition
(represented by the dashed line in Fig. 3). We can observe that
the TPA signal is enhanced by increasing the overlap between the
joint spectrum and the dashed line. Therefore, as one might
expect, the maximum overlap is obtained for a continuous wave
pump, with its frequency coinciding exactly with the two-photon
resonance condition.

\end{widetext}


\end{document}